\documentclass[aps,prb,showpacs,twocolumn]{revtex4}
\usepackage{color}
\usepackage[dvips]{graphicx}

\begin{document}

\title{
Neutron diffraction study of TbMnO$_3$: Magnetic structure revisited \\
}

\date{\today}

\author{R. Kajimoto}
\affiliation{Neutron Science Laboratory, Institute
of Materials Structure Science, High Energy Accelerator Research
Organization (KEK), 1-1 Oho, Tsukuba, Ibaraki 305-0801, Japan}

\author{H. Yoshizawa}
\affiliation{Neutron Science Laboratory, Institute for Solid State
Physics, University of Tokyo, Tokai, Ibaraki 319-1106, Japan}

\author{H. Shintani}
\affiliation{Department of Applied Physics, University of Tokyo,
Bunkyo-ku, Tokyo 113-8656, Japan}

\author{T. Kimura}
\altaffiliation[Present address: ]
{Los Alamos National Laboratory, Los Alamos, New Mexico 87545, USA.}
\affiliation{Department of Applied Physics, University of Tokyo,
Bunkyo-ku, Tokyo 113-8656, Japan}

\author{Y. Tokura}
\affiliation{Department of Applied Physics, University of Tokyo,
Bunkyo-ku, Tokyo 113-8656, Japan}

\begin{abstract}

 Magnetic ordering in TbMnO$_{3}$ was revisited by a neutron diffraction
 study. In addition to the previously reported \textit{A}-type and
 \textit{G}-type modulated structures with a propagation vector
 $(0,q_\mathrm{Mn},0)$,\cite{quezel77,blasco00} we found \textit{C}-type
 and \textit{F}-type orderings of the Mn moments. All the components
 appear below $T_N^\mathrm{Mn} = 46$~K, and the \textit{G}, \textit{C},
 and \textit{F}-type components are enhanced below $T_\mathrm{lock} =
 28$~K, where $q_\mathrm{Mn}$ ordering locks at its low temperature
 value. The locking of the propagation vector also yields the squaring
 up of the spin arrangement. The magnetic moments of Tb ions show a
 quasi long-range ordering below $T_N^\mathrm{Tb}=7$~K. It drastically
 promotes the development of the \textit{G}, \textit{C}, and
 \textit{F}-type components while suppressing the \textit{A}-type
 components.

\end{abstract}


\maketitle


Perovskite manganese oxides $R$MnO$_3$ ($R = \mbox{trivalent rare earth
ion}$) are famous as the parent materials of the colossal
magnetoresistive manganites. The Mn ions in $R$MnO$_{3}$ become
trivalent having the $t_{2g}^3 e_{g}^1$ configuration. This electronic
configuration induces a degree of freedom of the $e_g$ orbital, which
produces many novel properties. In the most well studied material,
LaMnO$_{3}$, the $e_g$ orbitals form staggered ordering of
$d_{3x^2-r^2}$ and $d_{3y^2-r^2}$ orbitals, which induces the
layered-type (so-called \textit{A}-type) antiferromagnetic ordering of
Mn spins. In contrast, when the ionic radius of $R$ is substantially
small, $R$MnO$_{3}$ shows a modulated spin ordering with a propagation
vector $(0,q_\mathrm{Mn},0)$ (in the orthorhombic \textit{Pbnm} cell).
\cite{quezel77,blasco00,munoz01,brinks01,munoz02} The emergence of the
incommensurate spin structure is explained as a competition between the
nearest-neighbor (NN) spin interactions and the next-nearest-neighbor
(NNN) interactions caused by the combination of the GdFeO$_{3}$-type
distortion and the $d_{3x^2-r^2}/d_{3y^2-r^2}$ orbital ordering (We will
refer to this model as the NN-NNN model.).\cite{kimura_prb}

TbMnO$_{3}$ is one of the latter series of $R$MnO$_{3}$. It shows an
incommensurate sinusoidal spin ordering with $q_\mathrm{Mn}\sim 0.295$
below $T_N \sim 41$~K with spins oriented along the [010]
direction. $q_\mathrm{Mn}$ decreases as temperature is lowered, then
locked at $q_\mathrm{Mn} = 0.28$ below $T_\mathrm{lock} \sim
30$~K.\cite{quezel77} Very recently, it was found that the spontaneous
electric polarization $P$ parallel to the $c$ axis appears below
$T_\mathrm{lock}$.\cite{kimura_nature} Moreover, the magnitude or the
direction of $P$ can be drastically changed by applying a magnetic
field.\cite{kimura_nature} The magnetic control of the ferroelectric
polarization will provide an attractive possibility of new
magnetoelectric devices.

These recent discoveries in the $R$MnO$_{3}$ series motivated us to
reinvestigate the magnetic properties of these materials in
detail. Here, we report the result of a neutron diffraction study on the
elastic properties of the magnetic ordering in TbMnO$_{3}$. The study of
the inelastic properties by neutron scattering is underway, and will be
published elsewhere.\cite{kaji_unpub}


A single crystal of TbMnO$_{3}$ was grown by the floating zone method. A
detailed procedure of the sample preparation was described
elsewhere.\cite{kimura_nature} The neutron diffraction experiments were
performed using the triple axis spectrometer GPTAS installed at the
JRR-3M research reactor in JAERI, Tokai, Japan.  A neutron wave length
of $k_i = 2.57$ {\AA}$^{-1}$ was selected by the 002 reflection of a
pyrolytic graphite (PG) monochromator. The horizontal collimation of
40$^{\prime}$-40$^{\prime}$-40$^{\prime}$-blank (from monochromator to
sample) was used for most of the measurements, while the
20$^{\prime}$-20$^{\prime}$-20$^{\prime}$-blank collimation was utilized
for the investigation of temperature dependence of the wave vector of
the magnetic ordering.  Two PG filters were placed before the
monochromator and after the sample to suppress contaminations of
higher-order harmonics.  The crystal structure has the \textit{Pbnm}
symmetry, and the measurements were performed in the $(0kl)$ scattering
plane ($a^* = 1.08$~{\AA}$^{-1}$ and $c^* = 0.853$~{\AA}$^{-1}$). The
sample was mounted in an aluminum can filled with helium gas, and was
attached to the closed-cycle helium gas refrigerator.  All the
temperature dependence data were collected upon heating to avoid
uncertainty due to the hysteresis.



\begin{figure}
 \includegraphics[width=0.95\hsize]{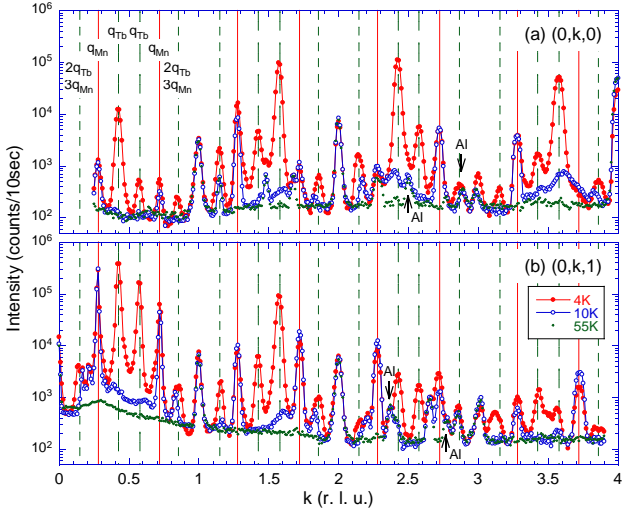}
 \caption{(Color online) Neutron diffraction profiles along (a) the
 $(0,k,0)$ line and (b) the $(0,k,1)$ line at 4, 10, and 55~K. Vertical
 lines indicate the positions of the Mn ($k\approx \mbox{integer} \pm
 2/7$ and 6/7) and Tb ($k\approx \mbox{integer} \pm 3/7$ and 6/7)
 superlattice peaks. ``Al'' denotes diffractions from the aluminum
 sample cell.}
 \label{kscan_profs}
\end{figure}

First, we surveyed magnetic reflections in the $(0kl)$ scattering plane
with the relaxed collimation
(40$^{\prime}$-40$^{\prime}$-40$^{\prime}$-blank). Figure
\ref{kscan_profs} shows neutron scattering profiles along the $(0,k,0)$
and $(0,k,1)$ lines at 4~K, 10~K, and 55~K. At 4~K, surprisingly many
superlattice reflections due to the magnetic moments of both Mn and Tb
ions are observed at every position with $k\approx n/7$
($n=\mbox{integers}$). At 10~K, on the other hand, only the Mn peaks are
observed with broad peaks coming from the short-range order of the Tb
moments.

The magnetic reflections due to the Mn ordering were observed at $(h,k
\pm q_\mathrm{Mn},l)$ with $q_\mathrm{Mn}=0.28$, where solid vertical
lines are drawn in Fig.~\ref{kscan_profs}. The observed Mn peaks can be
classified into five groups depending on the values of $h$, $k$, and $l$
as \textit{A}-type ($h+k=\mbox{even}$ and $l=\mbox{odd}$),
\textit{G}-type ($h+k=\mbox{odd}$ and $l=\mbox{odd}$), \textit{C}-type
($h+k=\mbox{odd}$ and $l=\mbox{even}$), and \textit{F}-type
($h+k=\mbox{even}$ and $l=\mbox{even}$). Quezel \textit{et al.}
observed strong \textit{A}-type reflections and weak \textit{G}-type
peaks, but they did not observe any \textit{C}-type and \textit{F}-type
reflections.\cite{quezel77} They concluded that the \textit{A}-type
peaks come from a sinusoidally modulated spin ordering with spins
parallel to the $b$ axis.  In contrast to Quezel's result, we observed
clear peaks of both \textit{C}-type and \textit{F}-type in addition to
the \textit{A}-type and \textit{G}-type peaks. Since the small number of
the observed magnetic peaks prevents us to unambiguously determine the
real space spin configuration, we just performed a preliminary analysis
of the magnetic structure: We calculated the intensities for the
sinusoidal and the helical spin configurations assuming the spins lie on
the $ab$, $bc$, or $ca$ planes, and compared them with the observed
intensities at 10~K. The intensities of the \textit{A}-type peaks are
best described by the sinusoidal model with spins parallel to the $b$
axis as reported by Quezel \textit{et al}. Analysis of the
\textit{G}-type peaks gave the similar result. For the \textit{C}-type
and \textit{F}-type peaks, however, we failed to satisfactorily
reproduce the observed intensities, though the existence of the peaks on
the $(0,k,0)$ line indicates the spin arrangements have substantial $x$
or $z$ components.

\begin{figure}
 \includegraphics[scale=0.65]{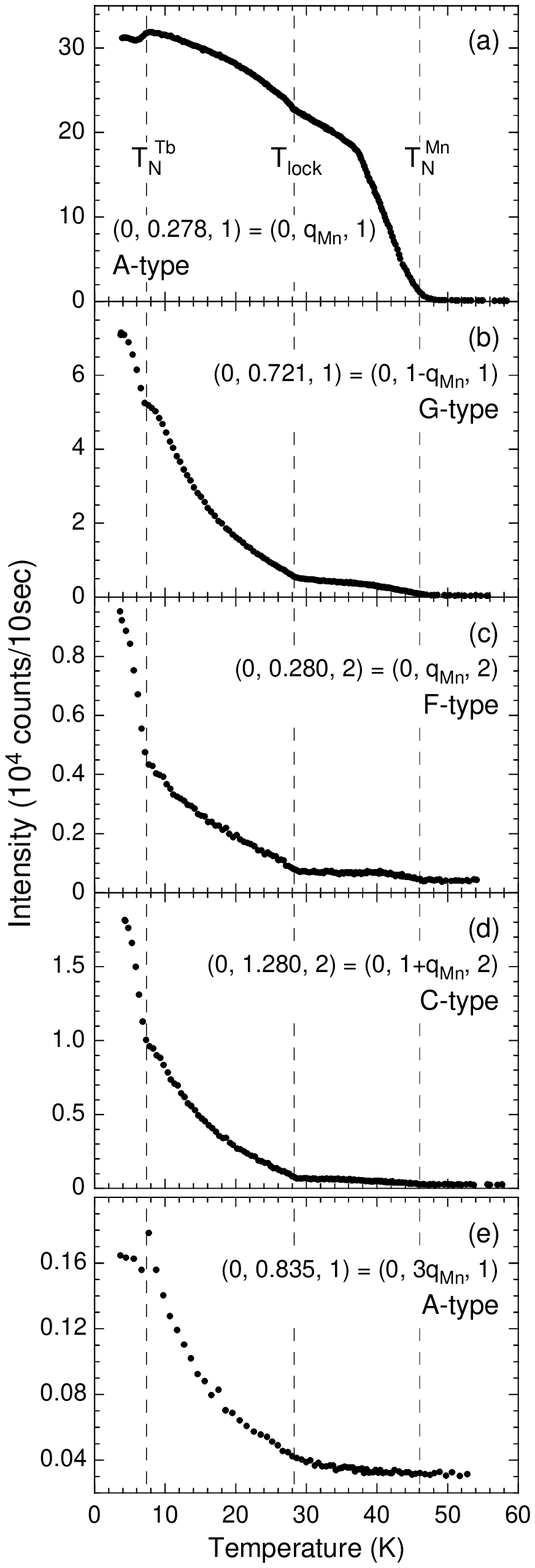} 
 \caption{(a)-(d) Temperature dependences of the intensities of the Mn
 superlattice peaks of (a) \textit{A}-type, (b) \textit{G}-type, (c)
 \textit{F}-type, and (d) \textit{C}-type. (e) Temperature dependence of
 the intensity of the Mn third harmonics at
 $(0,0.835,1)=(0,3q_\mathrm{Mn},1)$.}
 \label{Mn_Tdep}
\end{figure}

Figures \ref{Mn_Tdep}(a)-(d) show temperature dependences of the
scattering intensities of the Mn magnetic peaks. Each panel corresponds
to the superlattice peak of (a) \textit{A}-type:
$(0,0.278,1)=(0,q_\mathrm{Mn},1)$, (b) \textit{G}-type:
$(0,0.721,1)=(0,1-q_\mathrm{Mn},1)$, (c) \textit{F}-type:
$(0,0.280,2)=(0,q_\mathrm{Mn},2)$, and (d) \textit{C}-type:
$(0,1.280,2)=(0,1+q_\mathrm{Mn},2)$.  All the intensities start to
increase at $T_N^\mathrm{Mn} = 46$~K, which is slightly higher than the
reported transition temperatures.\cite{quezel77,blasco00,kimura_prb}
There are two distinct anomalies in the temperature profiles below
$T_N^\mathrm{Mn}$: with decreasing temperature, the slopes show upturns
at $T_\mathrm{lock} = 28$~K, and then a drastic drop of the
\textit{A}-type peak (a) or increases of the others (b)-(d) below
$T_N^\mathrm{Tb} = 7$~K. The former temperature coincides with the
temperature where the wave vector of the Mn ordering locks at its low
temperature value as described later. The latter anomaly is concomitant
with the ordering of the Tb moments (see below).  All the weak
(\textit{G}, \textit{C}, and \textit{F}) components show a similar
temperature dependence, and they gradually develop below
$T_N^\mathrm{Mn}$. Their temperature dependence is quite different from
the sharp increase of the \textit{A}-type component, suggesting that the
weak components arise from some secondary effect. One of probable causes
is the development of the ordering of the Tb moments, which is
manifested by the steep increase of these components in compensation for
the \textit{A}-type component on the (quasi) long-range ordering of the
Tb moments (Fig.~\ref{Mn_Tdep}).

In addition to the fundamental ($1q_\mathrm{Mn}$) peaks described above,
we also observed the third harmonics ($3q_\mathrm{Mn}$) of each type of
the Mn ordering at $(h,k \pm 3q_\mathrm{Mn},l)$ (dashed lines in
Fig.~\ref{kscan_profs}).  Though we could not detect the third harmonics
for the \textit{F}-type ordering, it may be because of the weakness of
the fundamental ($1q_\mathrm{Mn}$) \textit{F}-type peaks.  The existence
of the higher harmonics of the Mn ordering means that the spin ordering
is not an ideal sinusoidal wave.  The ratio of the intensity of the
$3q_\mathrm{Mn}$ peak to that of the $1q_\mathrm{Mn}$ peak is
$\sim$10$^{-2}$. This value is much smaller than that for an ideal
square wave (1/9), indicating the deviation of the spin arrangement from
the sinusoidal wave is quite small. Nevertheless, the existence of the
higher harmonics is interesting in that the NN-NNN model predicts the
square ordering of the Ising-like spins.\cite{kimura_prb} In
Fig.~\ref{Mn_Tdep}(e) is shown the temperature dependence of the
intensity of the $3q_\mathrm{Mn}$ peak of the \textit{A}-type
ordering. It gradually develops below $T_\mathrm{lock}$, indicating that
the sinusoidal ordering pattern of the Mn moments becomes distorted at
$T < T_\mathrm{lock}$.

\begin{figure}
 \includegraphics[scale=0.65]{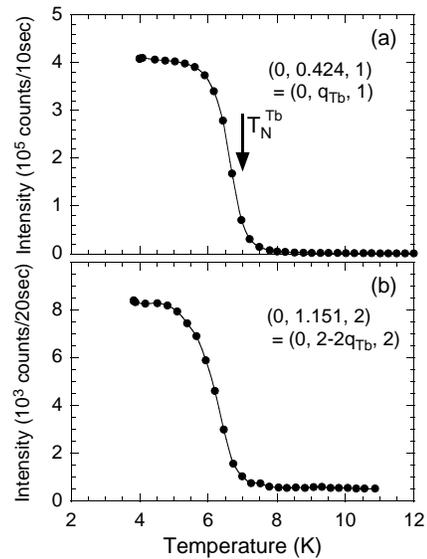} 
 \caption{Temperature dependences of the peak intensities due to the Tb
 ordering at (a) $(0,0.424,1) = (0,q_\mathrm{Tb},1)$ and (b)
 $(0,1.151,2) = (0,2-2q_\mathrm{Tb},2)$.}
 \label{Tb_Tdep}
\end{figure}

The superlattice peaks due to the ordering of the Tb moments were
observed at $(h,k \pm q_\mathrm{Tb},l)$ with $h,k,l=\mbox{integer}$ and
$q_\mathrm{Tb}=0.42$ (dotted-and-dashed lines in
Fig.~\ref{kscan_profs}). The Tb peaks are observed in all zones similar
to the Mn ordering. Though the previous study on a single crystal
reported the Tb peaks are broad and less intense compared to the Mn
peaks,\cite{quezel77} we observed quite sharp and much intense peaks at
the lowest temperature. Figure \ref{Tb_Tdep}(a) shows the temperature
dependence of the Tb peak intensity at
$(0,0.424,1)=(0,q_\mathrm{Tb},1)$. It shows a steep increase below
$T_N^\mathrm{Tb} = 7$~K.  In addition, superlattice peaks are observed
at $k \pm 0.15$ in all zones (dashed lines in
Fig.~\ref{kscan_profs}). The intensity of these peaks shows the same
temperature dependence as the Tb peaks [Fig.~\ref{Tb_Tdep}(b)],
indicating that these peaks are related to the Tb ordering. Since the
positions of the superlattices correspond to $2q_\mathrm{Tb}$, these
peaks may be attributed to the second harmonics of the Tb ordering. The
existence of the even harmonics means a space symmetry
breaking.\cite{chatt94} On the other hand, a magnetic ordering can
induce the second harmonics of lattice origin.\cite{pynn76} The exact
origin of the peaks at $2q_\mathrm{Tb}$ remains an open question.

Next, we measured temperature dependence of a profile of an
\textit{A}-type Mn peak to derive the change of the wave vector of the
Mn spin ordering as a function of temperature. The NN-NNN model predicts
that the wave number shows a step-wise temperature variation as locking
at certain rational values.\cite{kimura_prb} In the previous diffraction
measurements, however, only a smooth temperature dependence below
$T_\mathrm{lock}$ was observed,\cite{quezel77,kimura_prb} suggesting
that the effect of the locking is, if it exists, relatively weak. In the
present study, we collected the temperature dependence data at
substantially fine intervals of temperature ($\Delta T \sim 0.5$~K) with
the tight collimation (20$^{\prime}$-20$^{\prime}$-20$^{\prime}$-blank)
to detect small anomalies in the wave vector of the Mn ordering. With
this condition, the widths of the Mn peaks are still resolution limited
for both the [010] and the [001] directions, while the Tb peaks have
finite intrinsic widths.  The correlation lengths of the Tb ordering,
which is derived from the peak width of $(0,q_\mathrm{Tb},1)$ peak at
5~K by assuming a Gaussian line shape and deconvoluting the resolution,
is $\sim$140~{\AA} both along the [010] and [001] directions.

\begin{figure}
 \includegraphics[scale=0.5]{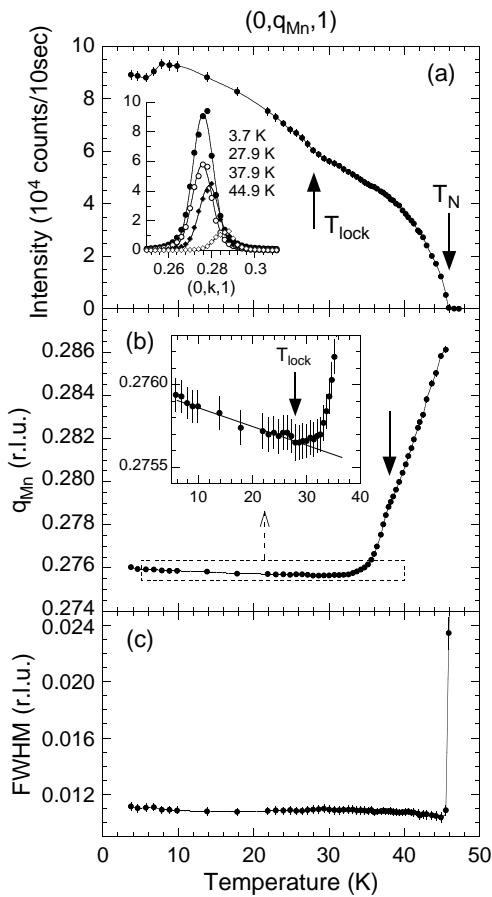}
 \caption{Temperature dependences of (a) the peak intensity, (b) the
 wave number, and (c) the peak width of the Mn $(0,q_\mathrm{Mn},1)$
 peak. Inset of panel (a): Temperature dependence of the Mn peak profile
 at $(0,q_\mathrm{Mn},1)$ along the [010] direction.}
 \label{Mn_prof_Tdep}
\end{figure}

We show in the inset of Fig.~\ref{Mn_prof_Tdep}(a) the temperature
variation of the profile of the Mn peak at $(0,q_\mathrm{Mn},1) \approx
(0,0.28,1)$. One can see clearly the peak position shifts toward higher
$Q$ as the temperature increases. By fitting these data to Gaussians
(solid lines in the figure), we extract the temperature dependence of
the peak height, the peak position (the wave number $q_\mathrm{Mn}$),
and the peak width (full width at half-maximum, FWHM), which are
summarized in Fig.~\ref{Mn_prof_Tdep}.  With increasing temperature, the
peak intensity shows an increase at $T < T_N^\mathrm{Tb}=7$~K, then
decreases [Fig.~\ref{Mn_prof_Tdep}(a)]. The slope of the temperature
profile exhibits an inflection at $T_\mathrm{lock}=28$~K. Finally, the
intensity vanishes at $T_N^\mathrm{Mn}=46$~K. For the wave number
$q_\mathrm{Mn}$, it is nearly constant at
$0.276~\mbox{r.l.u. (reciprocal lattice unit)} \approx
5/18$ (or 11/40)~r.l.u. at low temperature with a slow decrease as a
function of temperature [Fig.~\ref{Mn_prof_Tdep}(b)]. It shows a steep
rise above $\sim$34~K, but the upturn of the wave number starts at
$T_\mathrm{lock}$ [inset of Fig.~\ref{Mn_prof_Tdep}(b)]. It changes the
slope of its temperature dependence at 38~K, then it increases
monotonically without any locking behavior. For the peak width, it is
resolution limited below $T_N^\mathrm{Mn}$, and diverges above
$T_N^\mathrm{Mn}$ [Fig.~\ref{Mn_prof_Tdep}(c)]. We also measured
temperature dependence of the profile along the [001] direction, but did
not find any anomalies.

According to the NN-NNN model, the wave number of the Mn ordering
$q_\mathrm{Mn}$ decreases in a step-wise manner as lowering the
temperature, and reaches zero in the ground state.  However, the
step-wise behavior is indistinct and $q_\mathrm{Mn}$ locks at the finite
value below $T_\mathrm{lock}$ in the real material, although the
observed temperature dependence of $q_\mathrm{Mn}$ semiquantitatively
agrees with the model. This means some higher order effects than the NN
and NNN spin interactions that stabilize the modulated structure must be
considered. The coupling between the Mn moments and the Tb moments may
play a role, but it fails to explain the fact that the similar
incommensurate spin ordering and the locking behavior at the low
temperature were also observed in orthorhombic YMnO$_{3}$, where the
rare earth ion has no magnetic moment.\cite{munoz02} Further studies are
needed to solve this problem, and the study of the dynamics will be
helpful.


To sum up the present study, we have investigated the ordering of the Mn
and Tb moments in a single crystal of TbMnO$_{3}$ by neutron
diffraction.  The ordering process of the magnetic moments are
summarized as follows: At $T_N^\mathrm{Mn} = 46$~K, the Mn moments shows
a sinusoidally modulated \textit{A}-type ordering with a propagation
vector $(0,q_\mathrm{Mn},0)$, as reported by Quezel \textit{et
al}.\cite{quezel77} In addition, small \textit{G}, \textit{C}, and
\textit{F}-type with the same propagation vector
coexist. $q_\mathrm{Mn}$ is $\sim$0.29 at $T_N^\mathrm{Mn}$, and
smoothly decreases as the temperature is lowered showing a weak anomaly
at 38~K.  At $T_\mathrm{lock} = 28$~K, $q_\mathrm{Mn}$ locks at $0.276
\approx 5/18$ or 11/40. The locking of the wave vector amplifies the Mn
moment, especially enhances the \textit{G}, \textit{C}, and
\textit{F}-type components, and squares the spin arrangements. The Tb
moments order below $T_N^\mathrm{Tb} = 7$~K, which promotes the
development of the additional Mn components.


The authors thank S. Ishihara and K. Hirota for valuable discussions and
critical reading of the manuscript. This work was supported by KAKENHI
from the MEXT, Japan.


\end{document}